\documentclass[aps,prd,twocolumn,superscriptaddress,nofootinbib,floatfix]{revtex4-1}
\usepackage{amsmath}
\usepackage[dvips]{graphicx}
\usepackage[english]{babel}
\newcommand{\be}{\begin{equation}}
\newcommand{\ee}{\end{equation}}
\newcommand{\ben}{\begin{eqnarray}}
\newcommand{\een}{\end{eqnarray}}

\begin{document}
\title{Considerations on the thermal equilibrium between matter and the cosmic horizon}
\author{Jos\'{e} Pedro Mimoso \footnote{E-mail: jpmimoso@fc.ul.pt}}
\affiliation{Faculdade de Ci\^{e}ncias, Departamento de F\'{\i}sica
\& Instituto de Astrof\'{\i}sica e Ci\^encias do Espa\c co, Universidade de Lisboa,
Ed. C8, Campo Grande, 1749-016 Lisboa,
Portugal}
\author{Diego Pav\'{o}n\footnote{E-mail: diego.pavon@uab.es}}
\affiliation{Departamento de F\'{\i}sica, Universidad Aut\'{o}noma
de Barcelona, 08193 Bellaterra (Barcelona), Spain.}
%%%%%%%%%%%%%%%%%%%%%%%%%%%%%%%%%%%%%%%%%%%%%%%%%%%%%%%%%%%%%%%%%%%%%%%%%%%%%%%%
\begin{abstract}
A common feature in the thermodynamic analysis of homogeneous and
isotropic world models is the assumption that the temperature of
the fluids inside the cosmic horizon  (including dark energy)
coincides with the temperature of the latter, whether it be either
the event or the apparent horizon. We examine up to what extent
this assumption may be justified, given that these temperatures
evolve under different time-temperature laws. We argue that while
radiation cannot reach thermal equilibrium with the horizon,
nonrelativistic matter may, and dark energy might though only
approximately.
\end{abstract}
\maketitle
%%%%%%%%%%%%%%%%%%%%%%%%%%%%%%%%%%%%%%%%%%%%%%%%%%%%%%%%%%%%%%%%%%%%%%%%%%%%%%%%

\section{Introduction}
\noindent Spatially homogeneous and isotropic universe models are
usually adopted as starting point in  cosmological studies, both
mathematical and observational. This class of models is %are
characterized by the Friedmann-Lema\^{\i}tre-Robertson-Walker
(FLRW) metric,
%%%%%%%%%%%%%%%%%%%%%%%%%%%%%%%%%%%%%%%%%%%%%%%%%%%%%%%%%%%%%%%%%%
\begin{equation}
{\rm d}s^{2} = - c^{2} {\rm d}t^{2} \, + \, a^{2}(t) \frac{{\rm
d}r^{2}}{1 - k r^{2}}\, + \, a^{2}(t) r^{2} {\rm d}\Omega^{2} \, ,
 \label{flrw}
\end{equation}
%%%%%%%%%%%%%%%%%%%%%%%%%%%%%%%%%%%%%%%%%%%%%%%%%%%%%%%%%%%%%%%%%%
which relies on the cosmological principle
\cite{robertson1,robertson2,walker} whose validity, at large
scales, has not been contradicted \cite{clarkson} and thus far
looks rather robust \cite{stebbins,marco}. In the latter
expression the parameter $k=0,\pm 1$ discriminates the three types
of spatial curvatures and $\Omega$ is the unit two-sphere.
\\  \

%%%%%%%%%%%%%%%%%%%%%%%%%%%%%%%%%%%%%%%%%%%%%%%%%%%%%%%%%%%%%%%%%%%%%%%%%%%%%%%%%%%%
\noindent These spaces entail for each observer  an apparent
horizon \cite{Plebanski:2006sd} of radius
%%%%%%%%%%%%%%%%%%%%%%%%%%%%%%%%%%%%%%%%%%%%%%%%%%%%%%%%%%%%%%%%%%%%%%%%%%%%%%%%%%%
\begin{equation}
\tilde{r}_{h} = [(H/c)^{2} \, + \, k a^{-2}]^{-1/2}
\label{eq:hradius}
\end{equation}
%%%%%%%%%%%%%%%%%%%%%%%%%%%%%%%%%%%%%%%%%%%%%%%%%%%%%%%%%%%%%%%%%%%%%%%%%%%%%%%%
with $H = \dot{a}/a$ the Hubble factor. Since the observer has no
information about what is going on beyond the horizon, the latter
has an entropy, namely: $S_{h} = k_{B} \pi
\tilde{r}^{2}_{h}/\ell^{2}_{\rm pl}$  and a temperature $T_{h} =
\hbar H/(2\pi k_{B})$. Clearly for the spatially flat FLRW metric,
$k = 0$, the apparent horizon coincides with the Hubble horizon
whose radius simplifies to $c H^{-1}$.

Before going any further it
is expedient to recall the meaning of the apparent horizon. We
begin deriving its radius  \textemdash  see e.g. \cite{bak-rey,Cai:2006rs} for
details. To this end  we %by
recast the FLRW metric as
%%%%%%%%%%%%%%%%%%%%%%%%%%%%%%%%%%%%%%%%%%%%%%%%%%%%%%%%%%%%%%%%%
\begin{equation}
{\rm d}s^{2} = h_{ab} {\rm d} x^{a}\, {\rm d} x^{b}\, + \, \tilde{
r}^{2}(x)\, {\rm d} \Omega^{2}\, , \label{eq:frwmetric2}
\end{equation}
%%%%%%%%%%%%%%%%%%%%%%%%%%%%%%%%%%%%%%%%%%%%%%%%%%%%%%%%%%%%%%%%%
where $x^{0} = ct$, $x^{1} = r$ and $h_{ab} = {\rm diag} \left[
-1, \frac{a^{2}}{1-k r^{2}}\right]$. The radius of the dynamical
apparent horizon is set by the condition $\, h^{ab}
\partial_{a} \tilde{r} \, \partial_{b} \tilde{r} =0$, where $\tilde{r} = a(t) r$. A
straightforward calculation produces Eq. (\ref{eq:hradius}). Note
that the expansion of the ingoing and outgoing null geodesic
congruences is given by
%%%%%%%%%%%%%%%%%%%%%%%%%%%%%%%%%%%%%%%%%%%%%%%%%%%%%%%%%%%%%%%%%%%%%%
\begin{equation}
\theta_{IN} = H - \frac{1}{\tilde{r}}\, \sqrt{1 \, - \, \frac{k
\tilde{r}^{2}}{a^{2}}} \; \, {\rm and} \; \, \theta_{OUT} = H +
\frac{1}{\tilde{r}}\, \sqrt{1 \, - \, \frac{k
\tilde{r}^{2}}{a^{2}}}\, , \label{eq:inout}
\end{equation}
%%%%%%%%%%%%%%%%%%%%%%%%%%%%%%%%%%%%%%%%%%%%%%%%%%%%%%%%%%%%%%%%%%%%%%%
respectively. A spherically symmetric spacetime region will be
called  ``trapped"  if the expansion of ingoing and outgoing null
geodesics,  normal to the spatial two-sphere of radius $\tilde{r}$
centered at the origin, is negative. By contrast, the region will
be called ``antitrapped" if the expansion of the geodesics is
positive. In normal regions outgoing null rays have positive
expansion and  ingoing null rays, negative expansion. Thus, the
antitrapped region is given by the condition $\tilde{r}
> (H^{2} + k a^{-2})^{-1/2}$. Clearly, the surface of the apparent
horizon is nothing but the boundary hypersurface of the  spacetime
antitrapped region. Obviously, in the case of an exact de Sitter
expansion, the apparent and event horizons coincide.
\\  \

\noindent When analyzing the thermodynamics of FLRW models the
hypothesis that the temperature of either matter or dark fluids
(or  both) inside the horizon equals the temperature of the latter
or, at least,  is proportional to it is often made \textemdash \
see e.g.
\cite{bousso,gmi,bw,setare1,yg,ninfa,karami,chakraborty,saha}.
This is more frequently seen in the case of dark
energy\footnote{Assuming it differs from the cosmological
constant, as the latter has neither entropy nor temperature.} for
there is no clue about  which expression its temperature  should
have. In the absence of further information, such a viewpoint
provides the most economical and simplifying assumption, although
no clear physical argument supports it.
\\  \

\noindent The purpose of this note is to discuss and clarify the
validity of the above hypothesis. While the issue of the validity
of the first and second laws of thermodynamics in cosmology has
attracted a great deal of interest
\cite{Cai:2006rs,Karami:2010zz,Chakraborty:2012cw,Mimoso:2013zhp,Harko:2015pma},
the present question has somewhat been overlooked. As it turns
out, neither radiation nor dark energy can be in thermal
equilibrium (equality of temperatures) with the horizon for a long
period of time. In the case of nonrelativistic matter, the
equilibrium is in principle attainable, and,  after being reached,
it can be stable forever. But  the latter does not hold in the
earliest stages of cosmic expansion. For simplicity, we restrict
ourselves to spatially flat universes.

%%%%%%%%%%%%%%%%%%%%%%%%%%%%%%%%%%%%%%%%%%%%%%%%%%%%%%%%%%%%%%%%%%%%%%%%%%%
\section{Relativistic matter}
\noindent A natural and necessary condition for the equilibrium
between  the horizon (say, Hubble horizon, or what amounts to the
same, the apparent horizon in a spatially flat universe) and
thermal radiation (e.g., black body photons) at temperature
$T_{\gamma}$ is that the wavelength of the photons at which the
Planck's spectrum peaks given by Wien's law,
%%%%%%%%%%%%%%%%%%%%%%%%%%%%%%%%%%%%%%%%%%%%%%%%%%%%%%%%%%%%%%%%%%%%%%%
\begin{equation}
2.82 = \frac{c \, h}{k_{B} \, \lambda_{m} \, T_{\gamma}}\, ,
\label{wienlaw}
\end{equation}
%%%%%%%%%%%%%%%%%%%%%%%%%%%%%%%%%%%%%%%%%%%%%%%%%%%%%%%%%%%%%%%%%%%%%%%
be no larger than the horizon radius,
%%%%%%%%%%%%%%%%%%%%%%%%%%%%%%%%%%%%%%%%%%%%%%%%%%%%%%%%%%%%%%%%%%%%%%
\begin{equation}
\lambda_{m} \leq c \, H^{-1} \, . \label{nolarger}
\end{equation}
%%%%%%%%%%%%%%%%%%%%%%%%%%%%%%%%%%%%%%%%%%%%%%%%%%%%%%%%%%%%%%%%%%%%%%%
\\  \

\noindent Using $T_{h} = \hbar \, H/(2 \pi \, k_{B})$ alongside
(\ref{wienlaw}) in (\ref{nolarger}) it follows that the ratio
between temperatures must fulfill
%%%%%%%%%%%%%%%%%%%%%%%%%%%%%%%%%%%%%%%%%%%%%%%%%%%%%%%%%%%%%%%%%
\begin{equation}
 \frac{T_{h}}{T_{\gamma}} \leq \frac{2.82}{4 \, \pi^{2}} < 1 \, .
\label{tempratio}
\end{equation}
%%%%%%%%%%%%%%%%%%%%%%%%%%%%%%%%%%%%%%%%%%%%%%%%%%%%%%%%%%%%%%%%%
Thus, condition (\ref{nolarger}) implies $T_{h} < T_{\gamma}$; no
thermodynamic equilibrium can prevail between the horizon and a
bath of black body photons inside it. Actually, nowadays the
temperature of the cosmic microwave radiation, $T_{CMB}$, exceeds
by about 31 one orders of magnitude $T_{h}$
\cite{deMartino:2015ema,Planck:2011ah}
%%%%%%%%%%%%%%%%%%%%%%%%%%%%%%%%%%%%%%%%%%%%%%%%%%%%%%%%%%%%%%%%%%
\begin{equation}
\frac{T_h}{T_{CMB}}\Big|_0 \le 10^{-31}\; .
\label{ordersofmagnitude}
\end{equation}
%%%%%%%%%%%%%%%%%%%%%%%%%%%%%%%%%%%%%%%%%%%%%%%%%%%%%%%%%%%%%%%%%%%%%
Nevertheless given the relationship
%%%%%%%%%%%%%%%%%%%%%%%%%%%%%%%%%%%%%%%%%%%%%%%%%%%%%%%%%%%%%%%
\begin{equation}
\left(\frac{T_{h}}{T_{\gamma}}\right)^{\cdot} = -
\frac{T_{h}}{T_{\gamma}}\, q \, H \, ,
\label{rateratiotemps1}
\end{equation}
%%%%%%%%%%%%%%%%%%%%%%%%%%%%%%%%%%%%%%%%%%%%%%%%%%%%%%%%%%%%%%%%%%
and the fact that the deceleration parameter  $q = -(1+\dot{H}
H^{-2})$ is negative at present, the gap between both temperatures
is slowly narrowing (because the current value of $H$ is very
small) though it will never vanish.

%%%%%%%%%%%%%%%%%%%%%%%%%%%%%%%%%%%%%%%%%%%%%%%%%%%%%%%%%%%%%%%%%%%
\section{Nonrelativistic matter}
\noindent To study whether thermal equilibrium between
nonrelativistic matter, of mass $m$ and temperature $T_{m}$, and
the horizon is feasible, we proceed  as follows. First, we bear in
mind that the energy density and pressure of this fluid are given
by $\rho = n \, mc^{2} + (3/2) n k_{B} T_{m}$ and $P = n k_{B}
T_{m}$, respectively, where $\, n \, $ is the number density of
massive particles. Then we explore whether the de Broglie
wavelength, $\lambda = h/p$, satisfies
the reasonable condition $\lambda < c H^{-1}$. \\
To do this recall that  $p = mv  \propto \sqrt{m T_{m}}$, thus the
heavier the particle species, the earlier the said condition is
satisfied.  Nevertheless, a simple estimate reveals that in
general this will occur after the decoupling of the corresponding
particles species.\\
On the other hand, within a very good approximation $p \propto
a^{-1}$; then if the universe is dominated by thermal matter, the
Hubble factor will be nearly given by $H \propto a^{-3/2}$, and,
consequently, $(1+z)^{1/2} < c$, where $z$ is the redshift. Thus,
at very early times the condition involving the de Broglie
wavelength will not be fulfilled, but eventually it will be and it will remain so
forever.
\\  \

\noindent If the universe is dominated by thermal matter and the
cosmological constant $\Lambda$, then, sooner or later, the Hubble
factor will very approximately obey  $H \propto \sqrt{\Lambda}$;
whereby $1+z < \Lambda^{-1/2}$. Accordingly, the de Broglie
condition will be met as soon as the cosmological constant starts
dominating the expansion, or even earlier.
\\  \

\noindent After having seen that, in principle, the equality
$T_{m} = T_{h}$ can be reached at some point during the cosmic
expansion, it is worthwhile to study  whether the equilibrium,
when it happens, will be stable. To do so we must consider the
total heat capacity of the system, horizon plus matter. The heat
capacity of the horizon is negative. Indeed keeping in mind that
$\partial H/\partial T_{h} > 0$ and that we have chosen $k = 0$,
it follows
%%}%%%%%%%%%%%%%%%%%%%%%%%%%%%%%%%%%%%%%%%%%%%%%%%%%%%%%%%%%%%%%%%%%%%%%%%%%%%%%%%%%%%%%%%%%%%%
\begin{equation}
C_{h} = T_{h} \frac{\partial S_{h}}{\partial T_{h}} = T_{h} \,
\frac{\partial \left(\frac{\pi k_{B} \, c^{2}}{\ell^{2}_{\rm pl}
H^{2}}\right)}{\partial H} \frac{\partial H}{\partial T_{h}} < 0
\, . \label{hc-horizon}
\end{equation}
%%%%%%%%%%%%%%%%%%%%%%%%%%%%%%%%%%%%%%%%%%%%%%%%%%%%%%%%%%%%%%%%%%%%%%%%%%
\\  \

\noindent On the other hand, the heat capacity at constant volume
of thermal matter inside the horizon is
%%%%%%%%%%%%%%%%%%%%%%%%%%%%%%%%%%%%%%%%%%%%%%%%%%%%%%%%%%%%%%%%%%%%%%%%%%%%%
\begin{equation}
C_{m} = \frac{4 \pi}{3}\tilde{r}^{3}_{h} \,
\frac{\partial}{\partial T_{m}}\left(n m  c^{2}+ \frac{3}{2} n
k_{B} T_{m}\right) = 2 \pi k_{B} n \frac{c^{3}}{H^{3}}\, .
\label{hcmatter}
\end{equation}
%%%%%%%%%%%%%%%%%%%%%%%%%%%%%%%%%%%%%%%%%%%%%%%%%%%%%%%%%%%%%%%%%%%%%%%%%%%%%
Stable thermal equilibrium between two systems whose respective
heat capacities are of opposite signs is possible only if their
combined heat capacity is negative \cite{ptl,werner}. In the case
at hand, the inequality $C_{h} + C_{m} < 0$ yields the lower bound
on the expansion rate
%%%%%%%%%%%%%%%%%%%%%%%%%%%%%%%%%%%%%%%%%%%%%%%%%%%%%%%%%%%%%%%%%
\begin{equation}
 H > n c \, \ell^{2}_{\rm pl}\, ,
\label{eq:lowerboundonH}
\end{equation}
%%%%%%%%%%%%%%%%%%%%%%%%%%%%%%%%%%%%%%%%%%%%%%%%%%%%%%%%%%%%%%%%%%%
i.e., $T_{h} > \hbar^{2} G n/(2 \pi k_{B} c^{2})$.  Recalling that
the de Broglie condition $\lambda = h/p < c H^{-1}$ is satisfied
at least at late times and that $ n \propto a^{-3}$, the
right-hand side of the above inequality involving the Planck's
length decreases faster than $H$  in most reasonable cosmological
models. Therefore, there is a long period (possibly lasting
forever) in which the equilibrium  \textemdash if it
occurs\textemdash between thermal matter and the apparent horizon
is stable.
\\  \

\noindent A somewhat related question is the current value of
$T_{m}$. This can be evaluated recalling that the latter was in
equilibrium with the cosmic microwave background radiation until
the redshift was about $10^{4}$. Using this, alongside the
relations $T_{\gamma} = T_{\gamma 0} (1+z)$ and $T_{m} = T_{m0}
(1+z)^{2}$ (although the latter expression is strictly valid for
pressureless matter only), yields $T_{m0} \sim 3 \times 10^{-4}$
K. Consequently, $T_{m}$ is at all times 4 orders of magnitude
closer to the temperature of the horizon than the cosmic photon
gas.
\\  \

\noindent In the case at hand, the rate of the ratio $T_{h}/T_{m}$
is readily found after realizing that $\dot{T}_{m}/T_{m} = - 2H$.
It results in
%%%%%%%%%%%%%%%%%%%%%%%%%%%%%%%%%%%%%%%%%%%%%%%%%%%%%%%%%%%%%%%%%%%%%%%%%%%%%%%%%%%%%%%%%%%%%%%%%%%%%%%%%%
\begin{equation}
\left(\frac{T_{h}}{T_{m}}\right)^{\cdot} = - \frac{T_{h}}{T_{m}}
(q \, - \, 1)\, H.
\label{rateratiotemps2}
\end{equation}
%%%%%%%%%%%%%%%%%%%%%%%%%%%%%%%%%%%%%%%%%%%%%%%%%%%%%%%%%%%%%%%%%%%%%%%%%%%%%%%%%%%%%%%%%%%%%%%%%%%%%%%%%%%%
Because of $q < 0$ nowadays, the gap $T_{m} - T_{h}$ is narrowing,
and  this happens faster than in the case of black body radiation, as given by
Eq. (\ref{rateratiotemps1}).
%%%%%%%%%%%%%%%%%%%%%%%%%%%%%%%%%%%%%%%%%%%%%%%%%%%%%%%%%%%%%%%%%%%%%%%%%%%%%%%%%%%%%%%
%%%%%%%%%%%%%%%%%%%%%%%%%%%%%%%%%%%%%%%%%%%%%%%%%%%%%%%%%%%%%%%%%%%%%%%%%%%%%%%%%%%%%%%
%%%%%%%%%%%%%%%%%%%%%%%%%%%%%%%%%%%%%%%%%%%%%%%%%%%%%%%%%%%%%%%%%%%%%%%%%%%%%%%%%%%%%%%

%%%%%%%%%%%%%%%%%%%%%%%%%%%%%%%%%%%%%%%%%%%%%%%%%%%%%%%%%%%%%%%%%%%%%%%%%%%%%%%%%%%%%%%
\section{Dark energy}
\noindent %Here we
We now explore whether dark energy, with an equation of
state $\, P = w \rho$, where $w \,$ is lower than $-1/3$ and not
necessarily constant, can be in stable thermal equilibrium with
the apparent horizon. First notice that, if dark energy is a
scalar field, it is unclear whether it may have an entropy and a
temperature. In principle, this would be possible if the scalar
field is  not in a pure quantum state, but rather in a  mixed
state. However, we do not know which case we are dealing with. In
the latter one, we can use the expression \cite{mauricio}
%%%%%%%%%%%%%%%%%%%%%%%%%%%%%%%%%%%%%%%%%%%%%%%%%%%%%%%%%%%%%%%%%%%%%%%%%%%%%%%%%%%%%%%%%%%
\begin{equation}
\frac{\dot{T}_{de}}{T_{de}} = - 3 H \left(\frac{\partial
P}{\partial \rho} \right)_{n}
\label{tevolde}
\end{equation}
%%%%%%%%%%%%%%%%%%%%%%%%%%%%%%%%%%%%%%%%%%%%%%%%%%%%%%%%%%%%%%%%%%%%%%%%%%%%%%%%%%%%%%%%%%%
to derive that $T_{de}$ grows with expansion, as the last factor
on the right-hand side is negative. Since $T_{h}$ decreases with
expansion, it is hard to see how thermal equilibrium between the
dark energy and the apparent horizon can prevail.
\\  \

\noindent From the above equation it is seen that the larger $H$,
the higher the rate of variation of  $T_{de}$. However, away from
the primeval inflationary era, in any sensible cosmological model
(e.g. no phantom dominated) $H$ always decreases implying that the
grow of $T_{de}$ goes steadily  down \textemdash modulo $w$ does
not increase.
\\  \

\noindent Also, in spite of that a state of stable thermal
equilibrium between dark energy and the apparent horizon is not
achievable, since both $H$ and its current time  variation are
extremely small, the rate of variation of their ratio
$T_{h}/T_{de}$ will also be tiny. Indeed,
%%%%%%%%%%%%%%%%%%%%%%%%%%%%%%%%%%%%%%%%%%%%%%%%%%%%%%%%%%%%%%%%%%%%%%%%%%%%%%%%%%%%%%%%%%%%%%%%%%%%%%%%%%
\begin{equation}
\left(\frac{T_{h}}{T_{de}}\right)^{\cdot} = - \frac{T_{h}}{T_{de}}
[1+q-3w]\, H \, .
\label{rateratiotemps3}
\end{equation}
%%%%%%%%%%%%%%%%%%%%%%%%%%%%%%%%%%%%%%%%%%%%%%%%%%%%%%%%%%%%%%%%%%%%%%%%%%%%%%%%%%%%%%%%%%%%%%%%%%%%%%%%%%%%
Then, given the present smallness of $H$, we may say that the
ratio between both temperatures stays practically constant during
most of the history of the universe. Furthermore,  the rate at
which both temperatures depart from each other currently is
decreasing.
\\  \

\noindent On the other hand, nowadays the size of the apparent
horizon changes little in a Hubble time (at most of order unity),
i.e.,
%%%%%%%%%%%%%%%%%%%%%%%%%%%%%%%%%%%%%%%%%%%%%%%%%%%%%%%%%%%%%%%%%%%%%%%%%%%%%%%%%%%%%%%%%%%%%%%%%%%%%%%%%%%%%
\begin{equation}
t_{H} \, \frac{\dot{\tilde{r}}_{h}}{\tilde{r}_{h}} = 1\, +\, q \,
, \label{changelittle}
\end{equation}
%%%%%%%%%%%%%%%%%%%%%%%%%%%%%%%%%%%%%%%%%%%%%%%%%%%%%%%%%%%%%%%%%%%%%%%%%%%%%%%%%%%%%%%%%%%%%%%%%%%%%%%%%%%%
where $t_{H} = H^{-1}$ and for simplicity we have assumed $k = 0$.
Therefore if both temperatures are similar at some not extremely remote
early time, they will also remain so  today, and  this will last for another
Hubble time at least.
% Therefore if both temperatures are similar at some not extremely
% early times, they will remain so also today and  for another
% Hubble time at least.

%%%%%%%%%%%%%%%%%%%%%%%%%%%%%%%%%%%%%%%%%%%%%%%%%%%%%%%%%%%%%%%%%%%%%%%%%%%%%%%%%%%%%%%%%%%%%%%%%%%%%%%%%%%
\section{Concluding remarks}
\noindent In summary, thermal equilibrium between radiation and
the cosmic horizon cannot be achieved because Wien's law yields a
wavelength larger than the horizon radius, Eq. (\ref{tempratio}),
at all times. Nonrelativistic particles can attain equilibrium at
some point in the expansion that depends on the particle mass. The
stability of this equilibrium is submitted to a condition that
sets a lower bound for the Hubble factor, Eq.
(\ref{eq:lowerboundonH}), but it is easily fulfilled. As for dark
energy (assuming the corresponding scalar field is not in a pure
quantum state), a stable equilibrium can never be accomplished;
but if at some point in the expansion $T_{h}$ and $T_{de}$ happen
to be equal or close to each other, then they will stay
practically so for most of the history of the universe. Thus the
hypothesis, made by several authors, of thermal equilibrium
between dark energy and the horizon is, in this regard, not
unjustified.

%%%%%%%%%%%%%%%%%%%%%%%%%%%%%%%%%%%%%%%%%%%%%%%%%%%%%%%%%%%%%%%%%%%%%%%%%%%%%%%%%%%%%%%%%%%%%%%%%%%%%%%%%
\section*{Acknowledgments}
The authors are indebted to Alberto Rozas for comments on an
earlier version of this manuscript. D.P. is deeply grateful to the
``Instituto de Astrof\'{\i}sica and Ci\^encias do Espa\c co (IA)",
where part of this work was done, for warm hospitality and
financial support. J.P.M. acknowledges the kind hospitality of the
Universitad Aut\'{o}noma de Barcelona during his visit. J.P.M. and
D.P. were supported by Funda\c{c}\~ao para a Ci\^encia e a
Tecnologia (FCT) through the research grant UID/FIS/04434/2013.
The authors also acknowledge the COST Action CA15117, supported by
COST (European Cooperation in Science and Technology). And J.P.M.
further acknowledges the Project 6818 FCT/DAAD/2016-17.

%%%%%%%%%%%%%%%%%%%%%%%%%%%%%%%%%%%%%%%%%%%%%%%%%%%%%%%%%%%%%%%%%%%%%%%%%%%%%%%%%%%%%%%%%%%%%%%%%%%%%%%
%%%%%%%%%%%%%%%%%%%%%%%%%%%%%%%%%%%%%%%%%%%%%%%%%%%%%%%%%%%%%%%%%%%%%%%%%%%%%%%%%%%%%%%%%%%%%%%%%%%%%%%
%%%%%%%%%%%%%%%%%%%%%%%%%%%%%%%%%%%%%%%%%%%%%%%%%%%%%%%%%%%%%%%%%%%%%%%%%%%%%%%%%%%%%%%%%%%%%%%%%%%%%%%
%%%%%%%%%%%%%%%%%%%%%%%%%%%%%%%%%%%%%%%%%%%%%%%%%%%%%%%%%%%%%%%%%%%%%%%%%%%%%%%%%%%%%%%%%%%%%%%%%%%%%%%
%%%%%%%%%%%%%%%%%%%%%%%%%%%%%%%%%%%%%%%%%%%%%%%%%%%%%%%%%%%%%%%%%%%%%%%%%%%%%%%%%%%%%%%%%%%%%%%%%%%%%%%
%%%%%%%%%%%%%%%%%%%%%%%%%%%%%%%%%%%%%%%%%%%%%%%%%%%%%%%%%%%%%%%%%%%%%%%%%%%%%%%%%%%%%%%%%%%%%%%%%%%%%%%

%%%%%%%%%%%%%%%%%%%%%%%%%%%%%%%%%%%%%%%%%%%%%%%%%%%%%%%%%%%%%%%%%%%%%%%%%
\end{document}